# BOOST: A fast approach to detecting gene-gene interactions in genome-wide case-control studies

Xiang Wan[1],[*] Can Yang[1],[*] Qiang Yang[2], Hong Xue[3], Xiaodan Fan[4], Nelson L.S. Tang[5] and Weichuan Yu[1],[†]

[1]Department of Electronic and Computer Engineering, The Hong Kong University of Science and Technology.
[2]Department of Computer Science and Engineering, The Hong Kong University of Science and Technology.
[3]Department of Biochemistry, The Hong Kong University of Science and Technology.
[4]Department of Statistics, The Chinese University of Hong Kong.
[5]Laboratory for Genetics of Disease Susceptibility, Li Ka Shing Institute of Health Sciences, The Chinese University of Hong Kong.

**Gene-gene interactions have long been recognized to be fundamentally important to understand genetic causes of complex disease traits. At present, identifying gene-gene interactions from genome-wide case-control studies is computationally and methodologically challenging. In this paper, we introduce a simple but powerful method, named 'BOolean Operation based Screening and Testing'(BOOST). To discover unknown gene-gene interactions that underlie complex diseases, BOOST allows examining all pairwise interactions in genome-wide case-control studies in a remarkably fast manner. We have carried out interaction analyses on seven data sets from the Wellcome Trust Case Control Consortium (WTCCC). Each analysis took less than 60 hours on a standard 3.0 GHz desktop with 4G memory running Windows XP system. The interaction patterns identified from the type 1 diabetes data set display significant difference from those identified from the rheumatoid arthritis data set, while both data sets share a very similar hit region in the WTCCC report. BOOST has also identified many undiscovered interactions between genes in the major histocompatibility complex (MHC) region in the type 1 diabetes data set. In the coming era of large-scale interaction mapping in genome-wide case-control studies, our method can serve as a computationally and statistically useful tool.**

Genome-wide case-control studies use high-throughput genotyping technologies to assay hundreds of thousands of single-nucleotide polymorphisms (SNPs) and relate them to clinical conditions or measurable traits. To understand underlying causes of complex disease traits, it is often necessary to consider joint genetic effects (Epistasis) across the whole genome. The concept of epistasis [2] was introduced around 100 years ago. It is generally defined as interactions among different genes. The existence of epistasis has been widely accepted as an important contributor to genetic variation in complex diseases such as asthma, cancer, diabetes, hypertension, and obesity [5]. As a matter of fact, most researchers believe that it is critical to model complex interactions to elucidate the joint genetic effects causing complex diseases. They have demonstrated the presence of gene-gene interactions in complex diseases, such as breast cancer [14] and coronary heart disease [13].

The problem of detecting gene-gene interactions in genome-wide case-control studies has attracted extensive research interest. The difficulty of this problem is the heavy computational burden. For example, in order to detect pairwise interactions from $500,000$ SNPs genotyped in thousands of samples, we need $1.25 \times 10^{11}$ statistical tests in total. A recent review [5] presented a detailed analysis on many popular methods, including PLINK [8], MDR [14], Tuning ReliefF [12], Random Jungle[1](i.e., Random Forest [3]) and BEAM [17]. Among them, PLINK was recommended as the most computationally feasible method that is able to detect gene-gene interactions in genome-wide data [2] [5]. PLINK finished pairwise interaction examination of 89,294 SNPs selected from the WTCCC Crohn's disease data set in 14 days.

Here we propose a new method, named 'BOolean Operation based Screening and Testing' (BOOST), to analyze all pairwise interactions in genome-wide SNP data. In our method, we use a boolean representation of genotype data which is designed to be CPU efficient for multi-locus analysis. Based on this data representation, we design a two-stage (screening and testing) search method. In the screening stage, we use a non-iterative method to approximate the likelihood ratio statistic in evaluating all pairs of SNPs and select those passing a specified threshold. Most non-significant interactions will be filtered out and the survival of significant interactions is guaranteed. In the testing stage, we employ the typical likelihood ratio test to measure the interaction effects of selected SNP pairs.

| Disease | BD | CAD | CD | HT | RA | T1D | T2D |
|---|---|---|---|---|---|---|---|
| $C^1$ | 10 | 16 | 8 | 7 | 350 | 4499 | 18 |
| $C^1$ & $C^2$ | 0 | 0 | 1 | 0 | 0 | 789 | 0 |
| $C^1$ & $C^2$ & $C^3$ | 0 | 0 | 1 | 0 | 0 | 91 | 0 |

Table 1: The number of interactions identified from seven diseases data set under different constraints. $C^1$ – significant threshold constraint: The significance threshold is 0.05 for the Bonferroni-corrected interaction $P$-value; $C^2$ – distance constraint: The physical distance between two interacting SNPs is at least 1Mb. This constraint is used to avoid interactions that might be attributed to the linkage disequilibrium (LD) effects [5]; $C^3$ – main effect constraint: The single-locus $P$-value should not be less than $10^{-6}$. This constraint is used to see whether there exist strong interactions without significant main effects because those SNPs with $P \geq 10^{-6}$ are usually filtered out in the typical single-locus scan.

## Results

We have applied BOOST to analyze data (14,000 cases in total and 3,000 shared controls) from the Wellcome Trust Case Control Consortium (WTCCC) on seven common human diseases: bipolar disorder (BD), coronary artery disease (CAD), Crohn's disease (CD), hypertension (HT), rheumatoid arthritis (RA), type 1 diabetes (T1D) and type 2 diabetes (T2D). The analysis of each disease data set with control samples took less than 60 hours (around 2.5 days) to completely evaluate all pairs of roughly $360,000$ SNPs [3] on a standard 3.0 GHz desktop with 4G memory running Windows XP system. The results under different constraints are reported in Table 1. For T1D, we discovered many gene-gene interactions in the MHC region (see detailed descriptions in the following section). For other six diseases, however, we did not find nontrivial interactions (except one interacting SNP pair in CD).

---

[*]These authors contributed equally to this work.
[†]corresponding author.
[1]http://randomjungle.com/

[2]Marchini et al. [10] demonstrated that it is feasible to test association allowing for interactions in genome-wide scale. Beside that, Random Jungle can handle genome-wide data efficiently. However, they aim at testing associations allowing for interactions, which is easier than testing interactions. Please check the supplementary document and [5] for detailed explanations of 'test of association allowing for interactions' and 'test of interactions'.

[3]$500,000$ SNPs are genotyped in $5,000$ samples, about $360,000$ SNPs can pass the quality control, see the supplementary for details.

**T1D & RA.** The MHC region in chromosome 6 has long been investigated as the most variable region in the human genome with respect to infection, inflammation, autoimmunity and transplant medicine [9]. The recent study conducted by WTCCC [15] has shown that both T1D and RA are strongly associated with the MHC region via single-locus association mapping. The top-left panel of Figure 1 shows that the single-locus association map does not reveal much difference between T1D and RA. In our study, BOOST reports 4499 interactions in the T1D data set (see Table 1), in which 4489 interactions (99.8%) are in the MHC region. As comparison, BOOST reports 350 interactions in the RA data set, in which 280 interactions (80.0%) are in the MHC region. Our genome-wide interaction map provides the evidence that the MHC region is associated with these two diseases in different ways. The bottom panel of Figure 1 gives detailed interaction maps in the MHC region for T1D and RA data. The LD map [4] of MHC region is provided in the top-right panel of Figure 1. These interaction maps, different from the LD map, reveal a distinct pattern difference between T1D and RA. Specifically, there are three subregions in the MHC region, namely, the MHC class I region (29.8Mb - 31.6Mb), the MHC class III region (31.6Mb - 32.3Mb) and the MHC class II region (32.3Mb - 33.4Mb). A closer inspection of the T1D interaction map indicates that strong interaction effects widely exist between genes within and cross three classes, while most significant interactions in RA only involve loci closely placed in the MHC class II region. The contrast of the interaction patterns between T1D and RA may explain their different aetiologies, which are not revealed by single-locus association mapping.

**Interactions without significant main effects detected in T1D.** The MHC region is a highly polymorphic region with a high gene density. Although previous reports [7, 15] using the single-locus scan have identified strong associations between MHC genes (such as HLA-DQB1 and HLA-DRB1) and T1D, it is still unclear which and how many loci within the MHC region determine T1D susceptibility. Interactions without significant main effects can provide additional information to help pinpoint disease-associated loci because SNPs involved in those interactions are usually filtered out in the single-locus scan. Among the selected 789 interacting pairs in T1D, 91 pairs have non-significant loci under the single-locus scan (all of them are listed in the supplementary). A careful inspection of these 91 interactions has identified two interesting interaction patterns between the MHC Class I and Class II. Figure 2 presents one interaction pattern between the region 31350k-31390k and the region 32810k-32860k in chromosome 6. Another pattern between the region 31350k-31390k and the region 32930k - 32960k in chromosome 6 is provided in the supplementary. The interactions between two regions in Figure 2 are listed in Table 2. All SNPs in these interactions display weak main effects while their joint effects are statistically significant. As Nejentsev et al. [7] argued that both the MHC class I and II genes should be considered to better understand type 1 diabetes susceptibility, our results further provide the evidence that the interaction effects between these two classes may contribute to the aetiology of type 1 diabetes.

### COMPUTATION TIME

From a practical point of view, a key issue of detecting gene-gene interactions in genome-wide case-control studies is the computational efficiency [5]. Cordell [5] reported that PLINK took about 14 days to test pairwise interactions of the selected 89,294 SNPs on a single node of a computer cluster. A rough estimation implies that it would take approximately 228 days[5] to test all pairwise interactions of 360,000 SNPs. Random Jungle took about 5 hours to handle the selected 89,294 SNPs genotyped in 5000 samples. It is unknown how much time Random Jungle will need to find interactions from 360,000 SNPs. However, Random Jungle aims at detecting association allowing for interactions rather than detecting interactions (see detailed explanations in the supplementary). Besides, Random Jungle has the difficulty of finding interacting SNP pairs displaying weak main effects because trees built in Random Jungle are conditional on the main effects of SNPs. BEAM took about 8 days to handle 47,727 SNPs using $5 \times 10^7$ Markov chain Monte Carlo iterations. Currently, BEAM fails to handle $500,000$ to $1,000,000$ SNPs genotyped in 5000 or more samples. Cordell [5] recommended PLINK and Random Jungle as two most computationally feasible methods.

---
[4] We calculate composite LD using the method by Zaykin et al. [16].
[5] Given the number of samples, the running time of PLINK is proportional to the square of the number of SNPs. Therefore, the rough estimation is calculated by $14 \times (360,000/89,294)^2 \approx 228$.

| SNP 1 | | SNP 2 | | Interaction |
|---|---|---|---|---|
| SNP | Single-locus $P$-value | SNP | Single-locus $P$-value | BOOST $P$-value |
| rs2524057 | $4.807 \times 10^{-1}$ | rs9276448 | $8.878 \times 10^{-3}$ | $5.362 \times 10^{-14}$ |
| rs2524057 | $4.807 \times 10^{-1}$ | rs5014418 | $1.116 \times 10^{-2}$ | $2.738 \times 10^{-13}$ |
| rs2853934 | $8.336 \times 10^{-2}$ | rs9276448 | $8.878 \times 10^{-3}$ | $2.507 \times 10^{-13}$ |
| rs2524115 | $1.215 \times 10^{-1}$ | rs9276448 | $8.878 \times 10^{-3}$ | $6.456 \times 10^{-13}$ |
| rs3873385 | $3.368 \times 10^{-1}$ | rs9276448 | $8.878 \times 10^{-3}$ | $3.186 \times 10^{-14}$ |
| rs3873385 | $3.368 \times 10^{-1}$ | rs5014418 | $1.116 \times 10^{-2}$ | $3.841 \times 10^{-14}$ |
| rs3873385 | $3.368 \times 10^{-1}$ | rs6919798 | $6.077 \times 10^{-2}$ | $4.257 \times 10^{-13}$ |
| rs396038 | $9.939 \times 10^{-2}$ | rs9276448 | $8.878 \times 10^{-3}$ | $5.894 \times 10^{-13}$ |

Table 2: The interaction SNP pairs in the two regions shown in Figure 2. The SNPs in the column 'SNP 1' reside in the gene HLA-B and The SNPs in the column 'SNP 2' locate at the block across the genes HLA-DQA2 and HLA-DQB2. They show strong interactions without displaying significant main effects.

Our method BOOST makes a tremendous progress. It evaluated all pairs of roughly 360,000 SNPs within 60 hours (around 2.5 days) on a standard desktop (3.0 GHz CPU with 4G memory running Windows XP professional x64 Edition system). The WTCCC phase 2 study will analyze over 60,000 samples of various diseases using either the Affymetrix v6.0 chip or the Illumina 660K chip. The shared control samples will increase from $3,000$ to $6,000$. Such an increase in numbers of SNPs and sample size are more demanding on the computation efficiency. We anticipate that BOOST is still applicable to analyze the new data sets.

### CONCLUSION
The large number of SNPs genotyped in genome-wide case-control studies poses a great computational challenge in identification of gene-gene interactions. During the last few years, there have been fast growing interests in developing and applying computational and statistical approaches to finding gene-gene interactions. However, many approaches fail to handle genome-wide data sets (e.g., $500,000$ SNPs and $5,000$ samples). This hampers identification of interactions in genome-wide case-control studies. In this paper, we present a method named 'BOOST' to address this problem. We have successfully applied our method to analyze seven data sets from WTCCC. Not only is BOOST computationally efficient, it also has the advantage over PLINK with respect to the statistical power (see simulation study in the supplementary). Our experiment results demonstrate that interaction mapping is both computationally and statistically feasible for hundreds of thousands of SNPs genotyped in thousands of samples.

### METHODS
**Boolean representation of genotype data.** Suppose we have $\mathcal{L}$ SNPs and $n$ samples. The data set is usually stored in an $\mathcal{L} \times n$ matrix. Each cell in this matrix takes a value in $\{1, 2, 3\}$, which represent homozygous reference genotype, heterozygous genotype and homozygous variant genotype, respectively. In our method, we introduce a Boolean representation of genotype data (the details are provided in the supplementary). This Boolean representation promotes not only the space efficiency but also the CPU efficiency because it only involves Boolean values and thus allows using fast logic (bitwise) operations to obtain contingency tables.

**Measuring interaction effects.** Logistic regression models are of-ten used to measure the strength of gene-gene interactions [4, 5]. Let $L_M$ and $L_F$ be the log likelihood of the logistic regression model $M_M$ with only main effect terms ($df = 4$) and the full model $M_F$ with both main effect terms and interaction terms ($df = 8$), respectively. An interaction effect is measured by the difference of the log likelihood evaluated at maximum likelihood estimation (MLE), i.e., $\hat{L}_F - \hat{L}_M$. The likelihood ratio statistic $2(\hat{L}_F - \hat{L}_M)$ is asymptotically $\chi^2$ distributed with $df = 4$. However, it is computationally unaffordable to directly use this measure to evaluate all pairs of SNPs in a genome-wide case-control study because there are hundreds of billions of pairs to be tested.

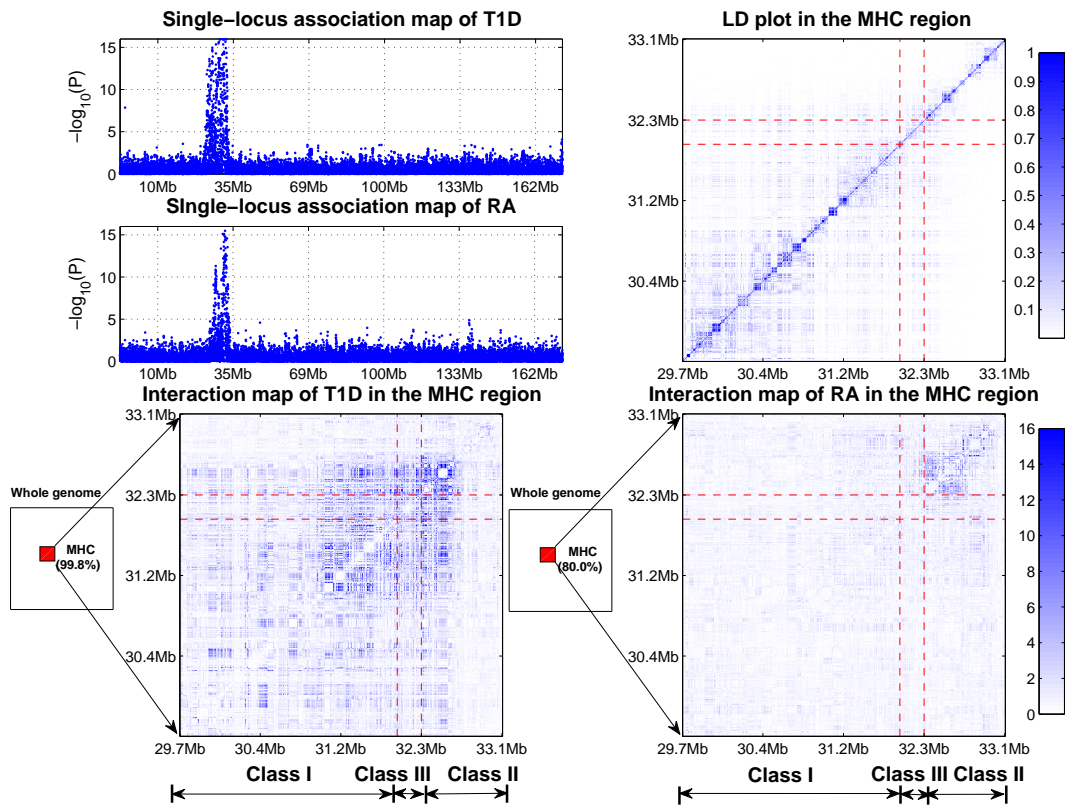

Figure 1: **Top-left panel:** Single-locus association mapping of T1D and RA. They share a very similar hit region in chromosome 6. **Top-right panel:** The LD map of the MHC region in control samples. **Bottom panel:** Genome-wide interaction mapping of T1D and RA. 99.8% interactions of T1D and 80.0% interactions of RA are in the MHC region. Strong interaction effects widely exist between genes in and across the MHC class I, II and III in T1D, while most significant interactions of RA only involve loci closely placed in the MHC class II region (The $P$ values are truncated at $P = 1.0 \times 10^{-16}$).

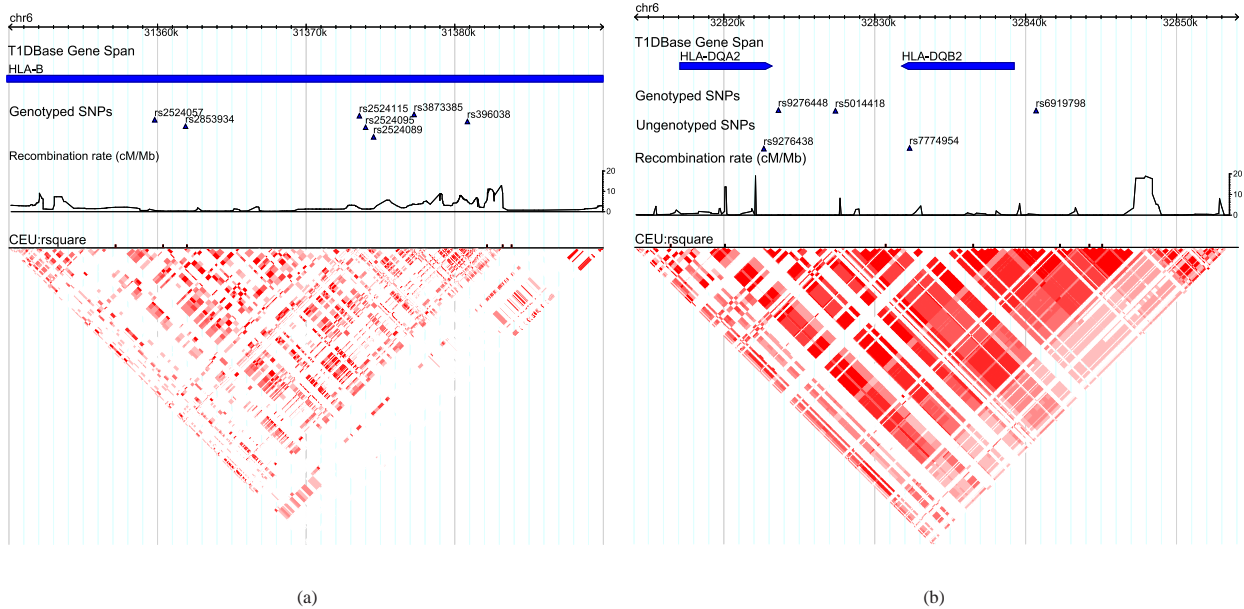

Figure 2: Interactions between the genes in the MHC class I and those in the MHC class II. **Left panel (a):** The region 31350k - 31390k of Chromosome 6. The gene HLA-B in the MHC class I locates in this region. The recombination rate and LD plot from HapMap show that a block structure spans from 31360k to 31380k. This region is mapped through the SNPs rs2524057, rs2853934, rs2524115, rs396038, rs3873385, rs2524095, and rs2524089. The SNPs rs2524095 and rs2524089 are involved in the interactions with the region 32930k - 32960k shown in Figure 1(b) of the supplementary. **Right panel (b):** The region 32810k - 32860k of Chromosome 6. The genes HLA-DQA2 and HLA-DQB2 in the MHC class II reside in this region. The recombination rate and LD plot from HapMap show that a block structure exists from 32820k to 32847k. This region is mapped through the genotyped SNPs rs9276448, rs5014418, and rs6919798. The ungenotyped SNPs rs9276438 and rs7774954 reside in the genes HLA-DQA2 and HLA-DQB2, respectively. They are in strong LD with those genotyped SNPs.

Noticing the correspondence between a logistic regression model and a log-linear model in categorical data analysis [1], we are able to measure an interaction effect under log-linear models. In the space of log-linear models, the homogenous model $M_H$ is the equivalent form of the main effect model $M_M$ and the saturated model $M_S$ is the equivalent form of the full model $M_F$. Let $\hat{L}_H$ and $\hat{L}_S$ be the log-likelihood of $M_H$ and $M_S$ evaluated at their MLEs, respectively. Interaction effects can thus be measured using $\hat{L}_S - \hat{L}_H$. After some algebra, it turns out that $\hat{L}_S - \hat{L}_H$ is connected with the Kullback-Leibler divergence [6]:

$$\hat{L}_S - \hat{L}_H = n \cdot D_{KL}(\hat{\boldsymbol{\pi}}||\hat{\mathbf{p}}), \quad (1)$$

where $n$ is the number of samples, $\hat{\boldsymbol{\pi}}$ is the joint distribution estimated under $M_S$ and $\hat{\mathbf{p}}$ is the joint distribution estimated under $M_H$. More details can be found in the supplementary.

**Boolean operation based screening and testing.** In Eq.(1), there is no closed-form solution for $\hat{\mathbf{p}}$ under the model $M_H$. Using iterative methods to estimate $\hat{\mathbf{p}}$ is computationally intensive to test all SNP pairs in genome-wide case-control studies. Here we use a non-iterative method known as 'Kirkwood Superposition Approximation' (KSA) [11] to approximate $\hat{\mathbf{p}}$, denoted as $\hat{\mathbf{p}}^K$. Let $\hat{L}_{KSA}$ be the log-likelihood of KSA model evaluated at its MLE. We show that

$$\hat{L}_S - \hat{L}_H \leq \hat{L}_S - \hat{L}_{KSA}. \quad (2)$$

In other words, $\hat{L}_S - \hat{L}_{KSA} = n \cdot D_{KL}(\hat{\boldsymbol{\pi}}||\hat{\mathbf{p}}^K)$ is an upper bound of $\hat{L}_S - \hat{L}_H$ (please check the supplementary for details). Based on this upper bound, we propose our new method 'BOolean Operation based Screening and Testing' (BOOST):

- Stage 1 (Screening): we evaluate all pairwise interactions by using KSA in the screening stage. For each pair, the calculation of $2(\hat{L}_S - \hat{L}_{KSA}) = 2n \cdot D_{KL}(\hat{\boldsymbol{\pi}}||\hat{\mathbf{p}}^K)$ is based on the contingency table collected by using Boolean operations. Since $2(\hat{L}_S - \hat{L}_H) \leq 2(\hat{L}_S - \hat{L}_{KSA})$, an interaction obtained by KSA without passing a specified threshold $\tau$, i.e., $2(\hat{L}_S - \hat{L}_{KSA}) \leq \tau$, would not be considered in Stage 2. We set the threshold $\tau = 30$ in our experiment.[6]

- Stage 2 (Testing): For each pair with $2(\hat{L}_S - \hat{L}_{KSA}) > \tau$, we test the interaction effect using the likelihood ratio statistic $2(\hat{L}_S - \hat{L}_H)$. We fit the log-linear models $M_H$ and $M_S$, and calculate this test statistic using Eq. (1). After that, we conduct the $\chi^2$ test with $df = 4$ to determine whether the interaction effect is significant.

---

[6]The threshold $\tau = 30$ corresponds to $P = 0.0002$ without Bonferroni correction, which is a very weak significant level for a genome-wide study. For the WTCCC data, the number of interactions passing this threshold ranges from $\sim 300,000$ to $\sim 600,000$.